# Selective-area chemical beam epitaxy of in-plane InAs one-dimensional channels grown on InP(001), InP(111)B, and InP(110) surfaces

Joon Sue Lee[1], Sukgeun Choi[2], Mihir Pendharkar[2], Dan J. Pennachio[3], Brian Markman[2], Micheal Seas[3], Sebastian Koelling[4], Marcel A. Verheijen[4], Lucas Casparis[5], Karl D. Petersson[5], Ivana Petkovic[5], Vanessa Schaller[6], Mark J.W. Rodwell[2], Charles M. Marcus[5], Peter Krogstrup[5], Leo P. Kouwenhoven[7], Erik P.A.M. Bakkers[4], and Chris J. Palmstrøm[1,2,3*]

[1]*California NanoSystems Institute, University of California, Santa Barbara, California 93106, USA*
[2]*Electrical and Computer Engineering Department, University of California, Santa Barbara California 93106, USA*
[3]*Materials Department, University of California, Santa Barbara California 93106, USA*
[4]*Department of Applied Physics, Eindhoven University of Technology, 5600 MB Eindhoven, The Netherlands.*
[5]*Center for Quantum Devices and Station-Q Copenhagen, Niels Bohr Institute, University of Copenhagen, 2100 Copenhagen, Denmark*
[6]*QuTech, Delft University of Technology, Delft, 2600 GA, The Netherlands*
[7]*Microsoft Station Q Delft, Delft, 2600 GA, The Netherlands*

ABSTRACT

We report on the selective-area chemical beam epitaxial growth of InAs in-plane, one-dimensional (1-D) channels using patterned $SiO_2$-coated InP(001), InP(111)B, and InP(110) substrates to establish a scalable platform for topological superconductor networks. Top-view scanning electron micrographs show excellent surface selectivity and dependence of major facet planes on the substrate orientations and ridge directions, and the ratios of the surface energies of the major facet planes were estimated. Detailed structural properties and defects in the InAs nanowires (NWs) were characterized by transmission electron microscopic analysis of cross-sections perpendicular to the NW ridge direction and along the NW ridge direction. Electrical transport properties of the InAs NWs were investigated using Hall bars, a field effect mobility device, a quantum dot, and an Aharonov-Bohm loop device, which reflect the strong spin-orbit interaction and phase-coherent transport characteristic in the selectively grown InAs systems. This study demonstrates that selective-area chemical beam epitaxy is a scalable approach to realize semiconductor 1-D channel networks with the excellent surface selectivity and this material system is suitable for quantum transport studies.

---

[*] Author to whom correspondence should be addressed. Electronic mail: cjpalm@ucsb.edu.

I. INTRODUCTION

There is an increasing interest in one-dimensional (1-D) III-V semiconductor nanowires (NWs) proximitized by superconductors that can potentially be used to realize topological superconducting phases [1,2]. Among III-V semiconductors, InAs and InSb have been of particular interest since InAs and InSb possess strong spin-orbit coupling and large g-factors. Indeed, signatures of Majorana zero modes have been experimentally observed from superconductor-paired InAs and InSb NWs prepared by vapor-liquid-solid (VLS) epitaxy method [3–6].

Recently, as a method of realizing the in-plane 1-D channels, selective-area epitaxy has been used and growths of In(Ga)As and InSb lateral NW networks have been demonstrated by molecular beam epitaxy and metal-organic chemical vapor phase epitaxy [7–13]. Since the network patterning and NW synthesis can be batch processed, this approach can be considered scalable. A superconducting layer such as Al can also be deposited *in vacuo* on the clean sidewall of NWs in a controllable manner. While the design and manipulation of NW networks are limited in the VLS method, selective-area epitaxy (SAE) offers greater flexibility.

Chemical beam epitaxy (CBE) is known to be a highly suitable growth technique for SAE of semiconductors on dielectric-patterned substrates [14]. Use of metal-organic group-III precursor molecules in CBE renders an excellent surface selectivity in comparison to conventional solid-source molecular beam epitaxy that has a narrow window of growth conditions to achieve both good surface selectivity and good film morphology. As a result, selective-area CBE has been used to monolithically integrate InP-based optoelectronic devices [14].

Here, we apply CBE to selectively grow in-plane InAs NWs on InP(001), InP(111)B, and InP(110) surfaces pre-patterned with $SiO_2$ layers. The dependence of the InAs morphology on the V/III ratio was first examined. Then, the optimum In-rich growth conditions were applied for SAE InAs NWs, exhibiting excellent surface selectivity. For the desired topological superconductivity studies, NWs with fewer defects and higher carrier mobility are required. We characterized structural and electrical properties of the InAs NWs as a function of substrate orientations and ridge directions. Furthermore, field effect mobility, spin-orbit interaction, electron g-factor, and phase coherence transport were investigated in the InAs NWs grown on InP(001) substrates.

II. EXPERIMENTAL

All the substrates used for SAE in this work are lithographically patterned, $SiO_2$-coated, Fe-doped semi-insulating InP(001), InP(111)B, and InP(110) wafers. A 3-nm-thick $Al_2O_3$ etch-stop layer was first deposited by atomic layer deposition on the bare wafers, followed by plasma-enhanced chemical vapor deposition of a 30-nm-thick $SiO_2$ masking layer. The patterns were defined by electron-beam lithography and formed by inductively coupled plasma etching with a $CHF_3/CF_4/O_2$ gas mixture. The $Al_2O_3$ etch-stop layer was finally etched by tetramethyl-ammonium hydroxide (TMAH – $N(CH_3)_4OH$). The InP surface was further cleaned by a sequence of ultraviolet-ozone cleaning for 900 seconds, diluted HCl ($HCl:H_2O=1:10$) etching for 60 seconds, and a final deionized $H_2O$ rinse for 30 seconds (The fabrication procedures are schematically shown in supplemental matarial S1.)

SAE was carried out in a VG-Semicon V80H CBE system. Trimethylindium (TMI – $In[CH_3]_3$) and thermally cracked arsine ($AsH_3$) and phosphine ($PH_3$) were used as source materials. A gas injector cell kept at 75°C was used for TMI and a modified low-pressure group-V hydride cracker cell kept at 1070°C was used for $AsH_3$ and $PH_3$. The growth chamber base pressure was $\leq 1\times10^{-10}$ mbar but increased as high as $\sim 2\times10^{-5}$ mbar during growth due to the large amount of $H_2$ created from the thermal decomposition of $AsH_3$ and $PH_3$ gases.

A piece ($\sim 7\times7$ $mm^2$) of reference InP(001) wafer was indium bonded onto the center of each 3-inch molybdenum sample holder for *in-situ* growth monitoring. Optical pyrometry and reflection high-energy electron diffraction (RHEED) were employed to measure the substrate temperature and characterize the surface during growth, respectively. The patterned substrates were cleaved into $\sim 5\times5$ $mm^2$-size square shapes, which were then co-mounted around the reference wafer on the same sample holder also using indium-bonding. In the VG-Semicon CBE system, the V/III ratios were controlled by regulating the pressure of the group-III and group-V precursor gas supply lines [15]. In general, the group-V line pressure was kept constant while altering the group-III line pressure.

Prior to growth, the native oxide was thermally desorbed from the substrate surface in a $P_2$ environment at ~520°C where the streaky (2×4) RHEED pattern is observed from the InP(001) reference sample. SAE was initiated by supplying TMI and growing a 9-nm-thick InP buffer layer. The line pressures for TMI and $PH_3$ were set to 0.5 Torr and 25 Torr, respectively. The resulting growth rate was ~0.5 monolayer (ML)/sec. Then, the TMI valve was closed and growth was interrupted with $PH_3$ for 30 seconds to smoothen the InP surface. InAs growth started by closing the $PH_3$ valve and opening the TMI and $AsH_3$ valves simultaneously. While the growth temperature was kept at 515~520°C, the V/III ratio was varied as a control parameter. To control the V/III ratio, the $AsH_3$ gas-line pressure was typically kept at 10 Torr and the TMI gas-line pressure varied. The TMI pressure was set to 0.3 Torr (~ 0.15 ML/sec) and 0.6 Torr (~ 0.60 ML/sec) for As-rich and In-rich growth, respectively. The transition from As-stabilized InAs(001) (2x4) RHEED patterns to In-stabilized InAs(001) (4x2) RHEED patterns was observed when the TMI pressure was at around 0.5 Torr.

For transport studies of topological superconductivity, it is crucial to have an electrically transparent interface between a semiconductor and a superconductor as evidenced in epitaxial Al-InSb NW devices which demonstrated the theoretically predicted quantized conductance [16,17]. We prepared the epitaxial Al in a separate ultrahigh-vacuum chamber interconnected to the CBE chamber. After the selective-area chemical beam epitaxy of the InAs NWs, we transferred the 3-inch molybdenum sample holder *in vacuo*. The epitaxial Al was grown by molecular beam epitaxy at a growth rate of 0.16 nm/minute with the sample holder cooled down to ~ 80 K by liquid nitrogen on a cold-head manipulator. A clean interface between the Al and InAs was observed in an InAs NW along [100] direction grown on an InP(001) substrate. Low-temperature transport measurements of tunneling spectroscopy of normal-superconductor junctions show a hard superconducting gap, but the superconductivity study is beyond the scope of this paper.

The morphology of InAs NWs, crystallographic orientation of major facets, and surface selectivity were examined by a FEI Nova Nano 650 FEG SEM. High-angle annular dark field scanning transmission electron microscopy (HAADF-STEM) using a probe-corrected JEOL ARM200F and a FEI Titan 300 kV TEM/STEM system was used for cross-sectional analysis of the NWs. The cross-sectional specimens were prepared using focused Ga ion beam (FIB) etching. Before FIB etching, a thin carbon layer was thermally evaporated on the surface of NWs followed by deposition of a 2.5-µm-thick Pt protection layer in the FIB. FIB etching steps down to 2 kV were utilized to minimize possible damage to the lamella. The TEM lamella was moved directly to an oxygen plasma cleaner before it was loaded into TEM/STEM systems to minimize contamination.

Electrical transport properties of the InAs NWs were characterized by Hall measurements at 2 K using a Quantum Design Physical Property Measurement System (PPMS) with magnetic fields up to 14 T. A standard lock-in technique was employed to measure the longitudinal voltage and Hall voltage in the Hall-bar geometry with an AC excitation current of 100 nA and a frequency of 13 Hz. The Hall bar channels with lengths of 800–1000 nm and widths of 140–180 nm were chosen for the measurements. For contact wiring, no post-growth fabrication was performed. Instead, indium dots were placed and annealed at 200°C for a few minutes on the grown InAs contact pads connected to the Hall bar channels, and the indium dots and pins on a measurement sample mount were connected via Au wires.

Further transport studies of field effect mobility, spin-orbit interaction, electron g-factor, and phase coherence were conducted. E-beam lithography followed by in-situ RF milling of the semiconductor surface before evaporating Ti/Au contacts was used to make electrical contacts to the InAs. A global $Al_2O_3$ dielectric was deposited before the Ti/Au top gate electrodes were evaporated. All measurements were performed at a base temperature (~30 mK) of a cryogen-free dilution refrigerator using lock-in amplifiers at 23 Hz excitation frequency.

III. RESULTS AND DISCUSSION

In order to optimize the growth conditions for selective-area CBE of InAs on InP, the conditions for blanket growth were first studied. For conventional heteroepitaxial growth of InAs on InP, it is well known [18] that the mechanisms of initial growth and surface roughening process for InAs on InP are influenced by the surface orientation and the V/III ratio. While InAs growth on GaAs(001) tends to exhibit a distinct transition from two-dimensional (2-D) layers to three-dimensional (3-D) islands (Stranski-Krastanov growth mode), the InAs growth on InP is characterized as a rather gradual changing from 2-D to 3-D surface morphology. Under In-rich condition, the transition is suppressed and 2-D InAs growth persists [19,20]. Schaffer *et al*. [21] found that InAs layers grown on GaAs(001) with In-rich condition exhibited improved structural and electrical properties. We grew InAs on InP(001) with a high V/III ratio, which led to rough 3-D islands. However, a low V/III ratio (slightly In-rich condition) resulted in much improved surface morphology. The P-stabilized InP(001) surface and In-stabilized InAs(001) surface were evidenced during the growth by (2x4) and (4x2) RHEED patterns, respectively. (See supplemental matarial S2 for RHEED images.) The streaky (4x2) pattern was continuously observed throughout InAs growth. InAs layers with As-rich condition typically show elongated spots in the [110] direction and chevron features in the [1-10] direction in the RHEED patterns. The surface morphology of 5.5-nm-thick InAs layers grown on InP(001) substrates, characterized using atomic force microscopy (AFM), are in good agreement with the previous results [22,23], which show that InAs initially forms nanometer-scale wire-like structures along the [1-10] direction with small 3-D islands at the kinks. The number of kinks and size of 3-D islands are found to increase with the V/III ratio, which results in a rough surface. (See supplemental matarial S3 for AFM results.) For InP and InAsP grown on InP(111)B, Hou and Tu [24] found that V/III incorporation ratio close to unity yielded a good surface morphology.

These optimized In-rich growth conditions for InAs blanket growth on InP(001) were then applied to selective-area CBE of InAs in-plane NWs on $SiO_2$-patterned InP surfaces. We found that InAs in-plane NWs form well-defined facets on InP(001) and InP(111)B surfaces under the In-rich growth conditions whereas on InP(110) surface well-faceted InAs NWs are seen under As-rich conditions.

A. Facets and surface energies

In Fig. 1(a), a top-view SEM image of InAs NWs grown on InP(001) substrate in the various ridge directions is shown. Starting from the [-110] direction, the NW patterns were prepared at every 15° in the counter-clockwise direction toward the [1-10] direction. This overview image reveals an excellent selectivity with no parasitic nucleation on the $SiO_2$ surface. Figures 1(b)–1(d) show the NWs with well-defined facets along the ridge orientations of [1-10], [110], [100], and [010], respectively. For the shape of the cross section, the NWs formed along the [1-10] direction possess a trapezoid shape with clear {111}A-plane side facets, confirmed by structural studies using TEM/STEM. The NWs along the [110] direction, in contrast, have almost rectangular cross-section with a small portion of {111}B-plane facets on the top edges. Although the relative portion of top-surface and side-wall facets varies with the growth conditions and ridge dimensions, the overall shapes of the NW cross-sections are in good agreement with the results reported by Krizek *et al*. [10]. The NWs grown along the <100> directions show the most consistent triangular cross-section with very smooth {110}-plane side facets, which was found less influenced by growth parameters.

Thermodynamics suggests that the facets of a crystal are formed such that the total surface energy is minimized. Below we calculate the ratios among the surface energies of different facets using a generic model of a trapezoidal crystal cross-section, as illustrated in Fig. 1(e). The area above the $SiO_2$ surface level is of interest for the calculation of the total surface energy per unit length along the NW, which is given by:

$$E_{total} = L_1\gamma_1 + L_2\gamma_2 + L_3\gamma_3 + L_4\gamma_4 + L_5\gamma_5, \qquad (1)$$

where $L_1, L_2, L_3, L_4$, and $L_5$ represent the lengths of facets in the cross section, and $\gamma_1, \gamma_2, \gamma_3, \gamma_4$ and $\gamma_5$ represent the surface energies of the facets.

For InAs NWs grown on an InP(001) substrate, the cross-sections of the NWs are symmetric ($L_1 = L_5$, $L_2 = L_4$, and $\theta_1 = \theta_2$ with $\theta_1$ and $\theta_2$ being angles between a plane parallel to the substrate and the facets of $L_2$ and $L_4$, respectively), and the Eq. (1) becomes

$$E_{total} = 2L_1\gamma_1 + 2L_2\gamma_2 + L_3\gamma_3, \qquad (2)$$

The cross-sectional area of the trapezoid as well as the width $w$ of a NW is set to be a constant.

$$Area = wL_1 + (w + L_3)(L_2 \sin\theta_1)/2 = constant \qquad (3)$$

$$w = 2L_2 \cos\theta_1 + L_3 = constant. \qquad (4)$$

The cross sections of the three types of NWs along [1-10], [110], and <100> directions on an InP(001) substrate are illustrated in Table I with the conditions of trapezoid side lengths, angle, and corresponding facet planes, obtained from the SEM studies in Fig. 1.

Substituting $L_1$ and $L_2$ in Eq. (2) with expressions of $L_3$ based on Eqs. (3) and (4) leads to the total surface energy $E_{total}$ per unit length along the NW as a quadratic function of a single variable $L_3$. By minimizing the quadratic function, the minimum $E_{total}$ is obtained with a condition where:

$$L_3 = w \left( \frac{\gamma_2 - \gamma_3 \cos\theta_1}{\gamma_1 \sin\theta_1} \right). \tag{5}$$

Using the experimental results and applying Eq. (5), the ratio between $\gamma_{(110)}$, $\gamma_{(111)A}$, $\gamma_{(100)}$, and $\gamma_{(111)B}$ can be estimated from different conditions of the three NW ridge directions:

$$\gamma_{(110)} : \gamma_{(111)A} : \gamma_{(100)} : \gamma_{(111)B} \cong 1.0 : 1.2 : 1.4 : 1.5. \tag{6}$$

Density functional theory (DFT) has been used to calculate the surface energies γ for (110), (100), (111)A, and (111)B planes of InAs grown on GaAs(001) substrate as a function of As chemical potential at the surface [25]. For In-rich growth condition, the surface energies for the InAs planes are given by:

$$\gamma_{(110)} < \gamma_{(111)A} < \gamma_{(100)} < \gamma_{(111)B}, \tag{7}$$

which is broadly in good agreement with the experimental results. The results from both our estimation and the DFT calculation by Moll *et al*. [25] suggest that (111)B facets are energetically unfavorable for InAs crystal grown with In-rich growth condition, which explains our observation of the small sections of (111)B facets formed on the NWs along the [110] ridge direction while the top (001) as well as the side-wall {110} facets are clearly seen. For the crystalline shape of NWs along the <100> directions, the slanted {110} planes have lower surface energy than the top (001) plane, which results in the triangular cross-section.

A top-view SEM image of InAs NWs grown on an InP(111)B surface is shown in Fig. 2(a). The NWs grown along the <11-2> directions exhibit rectangular cross-sections with the vertical {1-10} side-wall facets and flat (111)B top surface [Fig. 2(b)], which suggests the surface energy of the vertical {1-10} side-wall facets is the lowest for the side walls of the InAs NWs along the <11-2> directions, grown on InP(111)B substrate with In-rich condition. Formation of similar vertical side-wall facets was found from a recent InAs NWs grown on GaAs(111)B substrate along the <11-2> directions [11]. We note that Friedl *et al*. [11] attributed the formation of flat (111)B top surface (instead of two {113}-plane facets) of the InAs NW grown on GaAs(111)B surface to high $As_4$ flux used in their study. However, we also observed similar flat (111)B top surface in the In-rich growth conditions. For the NWs grown along the <1-10> ridge directions, we observed an asymmetric trapezoid cross-sections consisting of a slanted facet on one side, a vertical side-wall facet on the other side, and a flat (111)B top-surface facet as shown in Fig. 2(c).

While the slanted {100} facet is one of the low-energy low-index planes, the vertical side-wall facet appears to be a high-index {11-2} plane.

Similarly to the surface energy estimation for InAs NWs grown on InP(001) substrates, we calculate the ratio between the surface energy $\gamma_{(112)}$ of {11-2} planes and other surface energies obtained previously, based on the conditions for the two types of NWs along <1-10> and <11-2> directions grown on an InP(111)B substrate as detailed in Table I.

The total surface energy per unit length of the trapezoid crystals grown on an InP(111)B substrate is given by:

$$E_{total} = L_1\gamma_1 + L_3\gamma_3 + L_4\gamma_4 + L_5\gamma_5 \tag{8}$$

where $L_1 = L_4 \sin\theta_2 + L_5$ and $\gamma_1 = \gamma_5$. Here, the expression for area is the same as Eq. (3), and the width $w$ is give by:

$$w = L_3 + L_4 \cos\theta_2 = constant. \tag{9}$$

By substituting $L_4$ and $L_5$ based on the area and the width $w$ equations [Eqs. (3) and (9)], $E_{total}$ is expressed as a quadratic function of a single variable $L_3$, and $E_{total}$ is minimized when:

$$L_3 = w\left(\frac{-\gamma_3 + \gamma_4/\cos\theta_2 + \gamma_5 \tan\theta_2}{2\gamma_5 \tan\theta_2}\right). \tag{10}$$

The ratio between $\gamma_{(112)}$ and $\gamma_{(100)}$ can be estimated by applying the conditions for NWs along <1-10> direction, shown in Table I. Including $\gamma_{(112)}$ in the previous result of Eq. (6) gives:

$$\gamma_{(110)} : \gamma_{(111)A} : \gamma_{(100)} : \gamma_{(111)B} : \gamma_{(112)} \cong 1.0 : 1.2 : 1.4 : 1.5 : 2.4. \tag{11}$$

The surface energy of InAs {11-2} plane has not been reported, to the best of our knowledge, while Jacobi *et al.* [26] theoretically studied the surface energy of faceted GaAs (112) planes where the results suggested that the surface energy of (112) plane is just slightly higher than those of low-index (110), (111)A, and (111)B planes, for Ga-rich surfaces. In addition to a different material system and surface chemical potential, inclusion of strain energy possibly makes the surface energies of InAs surfaces on InP different from those of GaAs.

InAs NWs grown on InP(110) substrates form well-defined facets under As-rich growth conditions. The variation of the growth conditions from In-rich to As-rich condition could change the surface energy of each facet. Based on the DFT calculation by Moll *et al.* [25], the surface energy for (111)B surface strongly depends on the As chemical potential at the surface, and for the As-rich growth condition:

$$\gamma_{(111)B} < \gamma_{(110)} < \gamma_{(111)A} < \gamma_{(100)}. \tag{12}$$

We estimated the surface energies of the facets on the InAs NWs grown on InP(110) substrates by the methods used for the calculation of surface energies for the InAs NWs grown on InP(001) and InP(111)B substrates under the In-rich condition determined for (001) growth. With the shape and the facet information of the NWs grown on InP(110) substrates in Table I, we obtained:

$$\gamma_{(111)B} : \gamma_{(110)} : \gamma_{(111)A} : \gamma_{(112)} : \gamma_{(100)} \cong 0.8 : 1.0 : 1.2 : 1.2 : 1.3, \tag{13}$$

which is qualitatively consistent with the calculation for the low-index InAs facets for As-rich conditions by Moll *et al*. [25]. We note that the result of the ratio of $\gamma_{(110)} : \gamma_{(111)A}$ from the In-rich growth condition was also applied for the As-rich growth condition in Eq. (13), as the theoretical calculation show that $\gamma_{(110)}$ and $\gamma_{(111)A}$ values barely change as the As chemical potential varies [25]. However the (111)B surface energy depends dramatically on the As chemical potential, suggesting that dramatic reduction in the relative surface energy of (111)B for growth on (011) [Eq. (13)] compared to (001) and (111)B [Eqs. (6) and (11)] arises from the differences in effective As chemical potential on these surfaces despite the same incident As/In flux ratio. Information on facet formation and surface energy values for lateral InAs NW grown on InP(111)B and InP(110) substrates have been relatively rare. The results reported in this work may stimulate a systematic theoretical study of lateral heterogeneous InAs NW systems.

B. Structural characterization

Detailed structural properties of InAs NWs grown on InP(001), InP(111)B, and InP(110) substrates were characterized by TEM/STEM. Cross-sections perpendicular to the ridge direction (Fig. 4) as well as cross-sections along the ridge direction (Fig. 5) were prepared in the NWs along [100] direction grown on InP(001) substrates [Figs. 4(a) – 4(c) and 5(a) – 5(e)], the NWs along [1-21] direction grown on InP(111)B substrates [Figs. 4(d), 4(e), and 5(f) – 5(k)], and the NWs along [-110] direction grown on InP(110) substrates [Figs. 4(f), 4(g), and 5(i) – 5(p)]. Due to the $\sim Z^2$ dependence of image contrast in HAADF-STEM with Z being the atomic number, the $SiO_2$ mask appears almost black relative to the InP substrates and InAs NW structures, as shown in Figs. 4(a) and 4(d).

As suggested by top-view SEM images shown in Fig. 1(d), an InAs NW along [100] direction grown on an InP(001) substrate has a triangular cross-section with {011}-plane side facets, confirmed by HAADF-STEM in Fig. 4(a). This InAs NW has a 7-nm-thick Al layer grown on it, and the interface between InAs and Al is atomically clean without an amorphous interfacial layer or secondary phases [Fig. 4(b)]. A higher magnification image of an interface between the InP(001) substrate and InAs(001) shows a continuous atomic lattice [Fig. 4(c)]. In Fig. 4(d), an InAs NW along <11-2> direction grown on an InP(111)B substrate has rectangular cross-section with only small deviations of the {-101} side planes resulting from the mask sidewalls. After the layer grows past the mask layer, the side-walls and top facet are well defined, with small {012} corner facets present. The interface between the InP(111)B substrate and InAs(111)B shows a well-defined atomic lattice [Fig. 4(e)]. Figures 4(f) and 4(g) show conventional TEM images of an InAs NW along [-110] direction grown on an InP(110) substrate. In the cross-section of the NW, stacking faults parallel to the {111} side-wall facets are clearly seen. In cubic III-V compound semiconductors, the low stacking fault energy results in the formation of the stacking faults on {111} planes [27]. Charge transport properties of the InAs NWs may be affected by the defects and the correlated strain fields, and thus studies of defects and disorder in the SAE NWs could be important for potential electronic device applications.

We further investigate the defects formed along the ridge direction of the InAs NWs grown on InP(001), InP(111)B, and InP(110) substrates. InAs NWs grown on InP(001) substrates have inclined {111} planes across the NWs as schematically illustrated in Fig. 5(a). Conventional TEM and HAADF-STEM of the cross-section along the [100] ridge direction confirms that the stacking faults form on the inclined {111} planes [Figs. 5(c) and 5(d)]. The white arrows, shown in Fig. 5(d), indicate the stacking fault planes. The stacking fault planes look broad because {111} planes are inclined from the zone axis of [010] direction and the TEM lamella has a finite thickness (~50 nm). We also observed misfit dislocations originated in the 3.3 % lattice mismatch between InP and InAs. A Fourier-filtered image from an HAADF-STEM image clearly shows a misfit dislocation at the interface [Fig. 5(e)].

For the NWs grown on InP(111)B substrates, the stacking faults form on the {111} planes that are parallel to the growth plane as schematically illustrated in Fig. 5(f). From a cross-section of an InAs NW along the [11-2] ridge direction, stacking faults parallel to the (111)B substrate plane as well as across the NW ridge direction were seen in a conventional TEM image in Fig. 5(h). HAADF-STEM images reveal multiple stacking faults in InAs near the interface between the InP substrate and the InAs layer [Figs. 5(i) and 5(j)]. The stacking faults across the NW ridge direction [Fig. 5(k)] are also formed on the {111} planes near the grain boundaries.

In the InAs NWs along [-110] direction grown on InP(110) substrates, stacking faults on the {111} planes are parallel to the ridge direction and the inclined side-wall facets [Fig. 5(l)]. The stacking faults may not be seen depending on where the NW is cut along the ridge direction, as in the HAADF-STEM image in Fig. 5(n), whereas multiple stacking faults were observed in a cross-section perpendicular to the ridge direction as shown in Fig. 4(f). STEM images of the interface between InP(110) and InAs(110) along the ridge direction [Figs. 5(o) and 5(p)] reveal misfit dislocations with a density of $3 \times 10^5$ cm$^{-1}$, indicating that the InAs layer grown under the As-rich condition is mostly relaxed on the InP(110) substrate. (The misfit dislocation density of fully relaxed InAs on InP is to be $5\times 10^5$ cm$^{-1}$.)

C. Electrical transport properties

As evidenced in the studies of the surface energy and structural characterization, facet formation of NWs shows a strong dependence on substrate orientation and ridge direction. An important follow-up question will be if the substrate orientation and the ridge direction affect the transport properties of NWs. To answer this question and thus to optimize the configuration of NW networks, we prepared InAs NW Hall bar structures on SiO$_2$-patterned InP(001), InP(111)B, and InP(110) substrates as shown in Fig. 6(a). The Hall bars were made along the [1-10], [110], and <100> directions on an InP(001) substrate, the <1-10> and <11-2> directions on an InP(111)B substrate, and the [001], [-110], and <111> directions on an InP(110) substrate. The InAs Hall bar channels are in the diffusive transport regime where the channel length (L = 800–1000 nm) is longer than the elastic mean-free path (20–100 nm) with the channel widths of 130–180 nm. Figures 6(b)–6(i) show that the resulting magnetoresistance (longitudinal channel resistance as a function of perpendicular magnetic field) and Hall resistance vary with substrate orientations and ridge directions. We note that the Hall resistance linearly depends on the magnetic field, and the fluctuations in the Hall resistance are from a mix of longitudinal resistance and Hall resistance. The magnetoresistance fluctuations, observed in all channels on InP(001), InP(111)B, InP(110) substrates, are reproducible and identical in both positive and negative magnetic fields, which are referred to as the universal conductance fluctuations [28]. Weak localization (WL) correction to magnetoconductance has been reported in self-assembled and etch-defined InAs NWs in the case where the phase coherence length $l_\phi$ is smaller than the spin-orbit interaction length $l_{so}$ [29–31]. However, we do not attribute the negative magnetoresistance to WL because of the large amplitude of the conductance fluctuations in the InAs Hall bar channels, measured at 2 K. In the following section, weak antilocalization (WAL) correction to magnetoconductance will be further discussed using an InAs NW with a top gate.

Carrier density $n$ and charge mobility $\mu$ were calculated based on the results of the longitudinal resistance and Hall measurements. In addition, the electron mean-free path $l_e$ was calculated based on the carrier density and the mobility by $l_e = e\hbar\mu k_F$, where $e$, $\hbar$, and $k_F$ are electron charge, reduced Planck constant, and the Fermi wave vector $k_F$, respectively. Table II summarizes carrier density, mobility, and mean-free path as a function of substrate orientation and NW ridge direction. Overall ranges of carrier density, mobility, and mean-free path are within 0.9–2.1 $\times 10^{17}$ cm$^{-3}$, 1000–5000 cm$^2$/V s, and 20–100 nm, respectively. Due to difference in carrier density, direct comparison of charge mobility in various NWs is not viable. Gate-voltage tuning of the carrier density will be necessary for the comparison. The values of the obtained mobility are comparable to earlier measurements on InAs NWs. We speculate that by changing the buffer layer material or thickness the mobility can be further increased.

### D. Phase-coherent quantum transport

Here, we further perform electrical transport characterization in SAE InAs NWs grown on InP(001) substrates. In order to obtain field effect mobility $\mu_{fe}$, conductance of an InAs NW structure along [1-10] direction with the width $w$ ~ 140 nm was measured as a function of gate voltage $V_G$ at ~30 mK. In the case of diffusive transport, the conductance can be written as $g(V_g) = \frac{\mu_{fe} C}{L^2}(V_g - V_{th})$, where $C$ is the gate capacitance, $L$ is the gated segment length, and $V_{th}$ is the threshold voltage. By defining the transconductance $g_t = \frac{\partial g}{\partial V_g}$, for known C and L, the field effect mobility can be determined. A linear fit to the steepest part of the pinch off curve yields the peak transconductance [red in Fig. 7(a)]. From $L$ ~ 3.4 μm and numerically simulated $C$ ~ 1.7 fF, we estimate a field effect mobility of ~2000 cm$^2$/V s, which is within the range of the charge mobility obtained by the Hall measurements.

Next, we focus on characterizing spin-orbit interaction in the SAE InAs by magnetotransport. The device shown in the Fig. 7(a) inset was measured in an out-of-plane magnetic field $B$. The differential magnetoconductance <$\Delta g$> is averaged over a gate voltage window of 80 mV to suppress universal conductance fluctuations and offset to be zero at $B = 0$. Figure 7(b) plots <$\Delta g$>$(B)$ at $V_g = 0$ V (purple) and $V_g = -1.2$ V (green). The WAL peak at $B = 0$ is clearly visible in both traces, indicating strong spin-orbit coupling over the whole gate-voltage range. To get a measure for the phase coherence length $l_\phi$ and the spin-orbit length $l_{SO}$, one can model the WAL correction to conductance. Different models can lead to different length scales, but in order to compare to literature values we follow the commonly used model which is valid in the diffusive limit ($l_{el} \ll w$). The magnetoconductance is expressed as

$$\Delta g(B) \propto \frac{3}{2}\left(\frac{1}{l_\phi^2} + \frac{1}{3l_{SO}^2} + \frac{1}{D\tau_B}\right)^{-\frac{1}{2}} - \frac{1}{2}\left(\frac{1}{l_\phi^2} + \frac{1}{D\tau_B}\right)^{-\frac{1}{2}}, \tag{14}$$

where the magnetic dephasing time is given by $\tau_B = \frac{Cl_m^4}{w^2 D}$ with $D$ being the diffusion constant and $l_m = \sqrt{\frac{\hbar}{eB}}$ being the magnetic length, which is on the order of the NW diameter for the fitting range ~0.18 T. The prefactor $C$ is geometry and system dependent. To allow for a comparison to previous work on InAs NWs we use $C = 3$. The black lines in Fig. 7(b) show the fits of Eq. (14) to the data using a fixed $w$. The extracted $l_{SO}$ and $l_\phi$ of ~140 nm and ~300 nm, respectively, are in agreement with previous experiments on InAs NWs [29–31]. As a function of gate voltage, both length scales $l_{SO}$ and $l_\phi$ decrease at negative gate voltage [Figs. 7(c) and 7(d)]. This reduction has been attributed to a suppression of spin-orbit coupling in the NWs and thus the disappearance of the WAL peak [30,31], which is consistent with the data in Fig. 7(b).

We now present measurements of the electron g-factor in the InAs structures along the [1-10] direction. The inset of Fig 8(a) shows an SEM image of a device that was fabricated on a wide InAs slab and has two split-gate electrodes operating symmetrically with a gate voltage $V_g$. When pinching off the two gates, sharp peaks in conductance were observed [Fig. 8(a)], which we interpret as the Coulomb blockade due to some accidentally formed quantum dot. We use bias spectroscopy to quantify the electron g-factor [32]. In Fig. 8(b) the differential conductance is plotted as a function of $V_g$ and the bias voltage $V_{bias}$ in an in-plane magnetic field $B = 1.6$ T, exhibiting typical Coulomb diamonds. Additional lines appear outside of the blockaded region at larger $V_{bias}$, indicating exited states. A cut at a constant $V_g$ (blue arrows) shows two distinct peaks as seen in Fig. 8(c). At a lower magnetic field, $B = 0.8$ T (orange), the peak splitting reduces, and at $B = 0$ T (green) there is only one single peak visible. We interpret the peak splitting $\Delta V$ as the Zeeman splitting of the quantum dot level with $\Delta V = g\mu_B B/e$, where $\mu_B$ is the Bohr magneton. Figure 8(d) plots $\Delta V$ for different fields. We find that for $B < 0.4$ T and $\Delta V < 80$ μV it is difficult to extract a splitting. We suspect that the quantum dot level is tunnel broadened, and thus the line width is not temperature limited. A linear fit to $\Delta V(B)$ [red line in Fig. 8(d)] yields an electron g-factor of 3.3. This value is consistent with a wide spread in g-factors measured in InAs quantum dots [33].

Finally, we investigate the phase-coherent transport based on the Aharonov-Bohm (AB) effect in an SAE InAs loop structure. Conductance through the loop threaded with flux Φ, given for various temperatures, was measured. The obtained oscillations have the period $h/e = \Phi \cdot S$, consistent with the measured area $S$ of the AB loop. Figure 9(a) shows the measured conductance at various temperatures between 30 mK and 700 mK. Slowly-varying background was subtracted from each curve, and the curves are offset for clarity. Each curve was then Fourier-transformed, and the fast Fourier transform (FFT) amplitude in Fig. 9(b) denotes the value of the integrated first-harmonic peak. The integration limits are given by the inner and outer ring circumference, as measured from an SEM image [Fig. 9(b) Inset]. Error bars stem from the uncertainty in measuring AB loop dimensions. The decay of the AB FFT amplitude, shown in Fig. 9(b), was then fitted to $a \cdot \exp(-l/l_\phi(T))$, where $l$ is the length of one arm of the loop, $l_\phi$ is the electron coherence length, and $a$ is a constant. We assume $l_\phi \propto T^{-1/2}$, the case for a diffusive transport. The dependence assuming a ballistic transport $l_\phi \propto T^{-1}$ was attempted but did not fit as well. The phase coherence length $l_\phi$ is found to be 2 µm or roughly the ring circumference at 50 mK. The resulting values of the phase coherence length from the AB effect differ from the phase coherence length from WAL, which is likely due to the fact that the dephasing process governing the two effects is different [34].

## IV. CONCLUSIONS

InAs in-plane NWs were selectively grown on SiO$_2$-patterned InP(001), InP(111)B, and InP(110) substrates by CBE with an excellent surface selectivity. At the substrate temperature of ~520°C, slight In-rich growth condition yielded smooth surface morphology of InAs NWs on InP(001) and InP(111)B surfaces whereas As-rich growth condition resulted in well-defined facets of InAs NWs on an InP(110) surface. Faceting of the InAs NWs largely depends on the surface orientation and the ridge direction of the InAs NWs. On the (001) surface, the NWs formed along the [1-10] direction possess a trapezoid cross-section with clear {111}A-plane side facets while the ones that formed along the [110] direction have almost rectangular cross-section with a small portion of {111}B-plane facets on the top edges. The NWs grown in between those two directions, along the [010] direction, show a triangular cross-section with smooth {110}-plane side facets. On (111)B surface, the NWs grown in the <1-10> direction show asymmetric trapezoid cross-section with slanted {100} and vertical {11-2} side-wall facets and a flat (111)B top surface, whereas the NWs grown in the <11-2> direction exhibit rather rectangular cross-section with vertical {1-10} facets and a flat (111)B top surface. On (110) surface, rectangular cross-sections are seen in the NWs grown in the [001] and <111> directions while an asymmetric trapezoid cross-section with slanted {111} side-wall facets and a flat (110) top surface is observed in the NWs grown in the [-110] direction.

The ratios of the surface energies of the observed facet surfaces were estimated based on the structural characterization, and the results are qualitatively consistent with theoretical predictions for both In-rich and As-rich growth conditions. STEM/TEM of cross-sections perpendicular to the ridge direction and cross-sections along the ridge direction further reveals the atomic structures of interfaces between the InAs NWs and InP(001), InP(111)B, and InP(110) substrates as well as defects in the NWs. At the interface a continuous atomic lattice was observed with misfit dislocations due to the lattice mismatch between InAs and InP. Multiple stacking faults of inclined {111} planes were observed along the ridge directions in the InAs NWs grown on an InP(001) substrate whereas stacking fault planes are parallel to the substrate in the NWs grown on an InP(111)B substrate. In the NWs along [-110] direction grown on an InP(110) substrate, stacking faults are parallel to both the ridge direction and the inclined {111} sidewall facets.

Electrical transport properties were characterized using selectively grown Hall bar channels on various ridge directions grown on InP(001), InP(111)B, and InP(110) substrates. Obtained carrier density and mobility of the InAs channels, measured at 2 K, are in the range of 0.9–2.1 × $10^{17}$ cm$^{-2}$ and 1000–5000 cm$^2$/V s, respectively. The mobility of the InAs NWs grown on InP(001) substrates was confirmed by field effect mobility. From WAL correction to the conductance using the field effect mobility device, spin-orbit length $l_{SO}$ and phase coherence length $l_\phi$ was estimated to be ~140 nm and ~300 nm, respectively. The WAL is suppressed by applying negative gate-voltage tuning, which corresponds to decreases in $l_{SO}$ and $l_\phi$. The electron g-factor in an InAs slab along [1-10] direction with two split-gate electrodes was estimated to be 3.3 based on the Zeeman splitting of a quantum dot level. Phase-coherent transport in an AB loop geometry revealed the phase coherence length $l_\phi$ to be ~ 2 μm at 50 mK, which suggests that the conductance-interferometry-based measurement schemes of braiding [35,36] could be viable in this SAE InAs systems.

## ACKNOWLEDGMENTS

This work was supported by Microsoft Project Q and the Danish National Research Foundation. Patterned substrates were prepared in UCSB Nano-fabrication Facility, and most of the structural characterization studies were performed at the UCSB MRL Shared Experimental Facilities (NSF DMR-1720256), a member of the NSF-funded Materials Research Facility Network. The patterning development was supported by NSF (NSF ECCS-1640030). C.M.M. ac knowledges support from the Villum Foundation. Solliance and the Dutch province of Noord-Brabant are acknowledged for funding the probe-corrected STEM facility.

TABLE I. Cross-sectional shapes and facets of the InAs NWs with substrate orientations and ridge directions. Lengths ($L_1$–$L_5$) of the facets and the angles ($\theta_1$ and $\theta_2$) of trapezoids are defined in the generic model illustrated in Fig. 1(e).

| Substrate | InP(001) | | | InP(111)B | | InP(110) | | |
|---|---|---|---|---|---|---|---|---|
| Channel direction | [1-10] | [110] | <100> | <1-10> | <11-2> | [001] | <111> | [-110] |
| Cross section | | | | | | | | |
| L | $L_1, L_5 = 0$ $L_2 = L_4$ $L_2 : L_3 \approx 1:1$ | $L_1 = L_5$ $L_2 = L_4$ $L_2 : L_3 \approx 1:9$ | $L_1, L_3, L_5 = 0$ $L_2 = L_4$ | $L_2, L_5 = 0$ $L_3 : L_4 \approx 1:1$ | $L_2, L_4 = 0$ $L_3 = w$ | $L_1 = L_5$ $L_2 = L_4$ $L_2 : L_3 \approx 1:9$ | $L_1 = L_5$ $L_2 = L_4$ $L_2 : L_3 \approx 1:9$ | $L_2 : L_3 : L_4$ $\approx 1:4:4$ $L_5 = 0$ |
| $\theta$ | $\theta_1 = 54.7°$ $\theta_2 = 54.7°$ | $\theta_1 = 54.7°$ $\theta_2 = 54.7°$ | $\theta_1 = 45°$ $\theta_2 = 45°$ | $\theta_1 = 0°$ $\theta_2 = 54.7°$ | $\theta_1 = 0°$ $\theta_2 = 0°$ | $\theta_1 = 45°$ $\theta_2 = 45°$ | $\theta_1 = 60°$ $\theta_2 = 60°$ | $\theta_1 = 35.3°$ $\theta_2 = 35.3°$ |
| Surface energy | $\gamma_2 = \gamma_{(111)A}$ $\gamma_3 = \gamma_{(100)}$ $\gamma_4 = \gamma_{(111)A}$ | $\gamma_1 = \gamma_{(110)}$ $\gamma_2 = \gamma_{(111)B}$ $\gamma_3 = \gamma_{(100)}$ $\gamma_4 = \gamma_{(111)B}$ $\gamma_5 = \gamma_{(110)}$ | $\gamma_2 = \gamma_{(110)}$ $\gamma_4 = \gamma_{(110)}$ | $\gamma_1 = \gamma_{(112)}$ $\gamma_3 = \gamma_{(111)B}$ $\gamma_4 = \gamma_{(100)}$ | $\gamma_1 = \gamma_{(110)}$ $\gamma_3 = \gamma_{(111)B}$ $\gamma_5 = \gamma_{(110)}$ | $\gamma_1 = \gamma_{(110)}$ $\gamma_2 = \gamma_{(100)}$ $\gamma_3 = \gamma_{(110)}$ $\gamma_4 = \gamma_{(100)}$ $\gamma_5 = \gamma_{(110)}$ | $\gamma_1 = \gamma_{(112)}$ $\gamma_2 = \gamma_{(110)}$ $\gamma_3 = \gamma_{(110)}$ $\gamma_4 = \gamma_{(110)}$ $\gamma_5 = \gamma_{(112)}$ | $\gamma_1 = \gamma_{(100)}$ $\gamma_2 = \gamma_{(111)A}$ $\gamma_3 = \gamma_{(110)}$ $\gamma_4 = \gamma_{(111)B}$ |

TABLE II. Transport properties of the InAs NWs of various substrate orientations and channel directions, obtained at 2 K. $n$, $\mu$, and $l_e$ represent carrier density, charge mobility, and mean-free path, respectively.

| Substrate | Channel directions | Channel width $w$ | Channel length $L$ | $n$ ($10^{17}$ cm$^{-3}$) | $\mu$ (cm$^2$/V s) | $l_e$ (nm) |
|---|---|---|---|---|---|---|
| InP(001) | [110] | 180 nm | 800 nm | 1.32 | 1170 | 22 |
| InP(001) | [1-10] | 170 nm | 800 nm | 1.43 | 1420 | 28 |
| InP(001) | <100> | 160 nm | 800 nm | 0.89 | 2450 | 38 |
| InP(111)B | <11-2> | 140 nm | 800 nm | 1.14 | 2430 | 43 |
| InP(111)B | <1-10> | 140 nm | 800 nm | 1.35 | 4760 | 91 |
| InP(110) | [001] | 130 nm | 1000 nm | 2.07 | 1040 | 22 |
| InP(110) | <111> | 130 nm | 1000 nm | 1.04 | 1520 | 23 |
| InP(110) | [-110] | 130 nm | 1000 nm | 1.55 | 1840 | 34 |

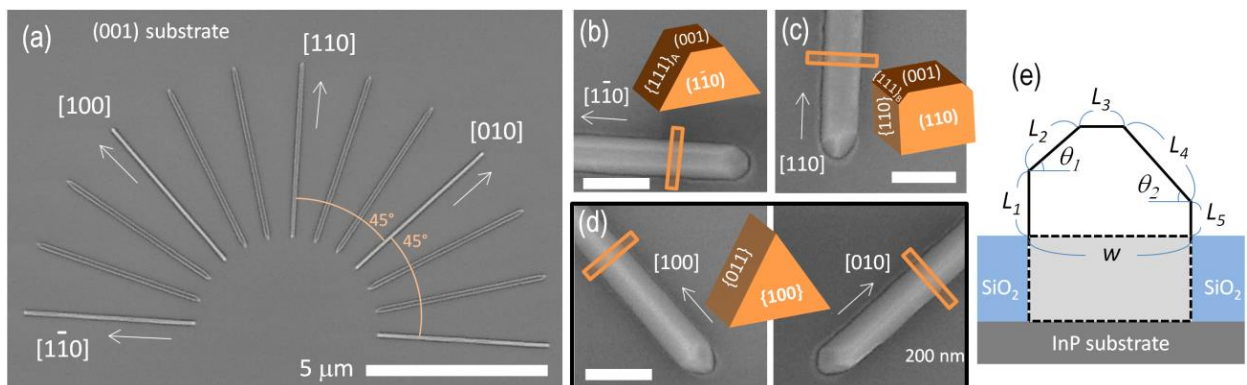

FIG 1. InAs NWs grown on an InP(001) substrate. (a) An SEM image of InAs NWs grown on an InP(001) substrate in various ridge directions. Well-defined facets are formed in particular ridge directions: (b) [1-10], (c) [110], and (d) <100>. Scale bars: 200 nm. (e) Schematic of a generic model of a trapezoidal crystal cross-section, applied to all InAs NWs grown on InP(001), InP(111)B, and InP(110) substrates for the surface energy calculation.

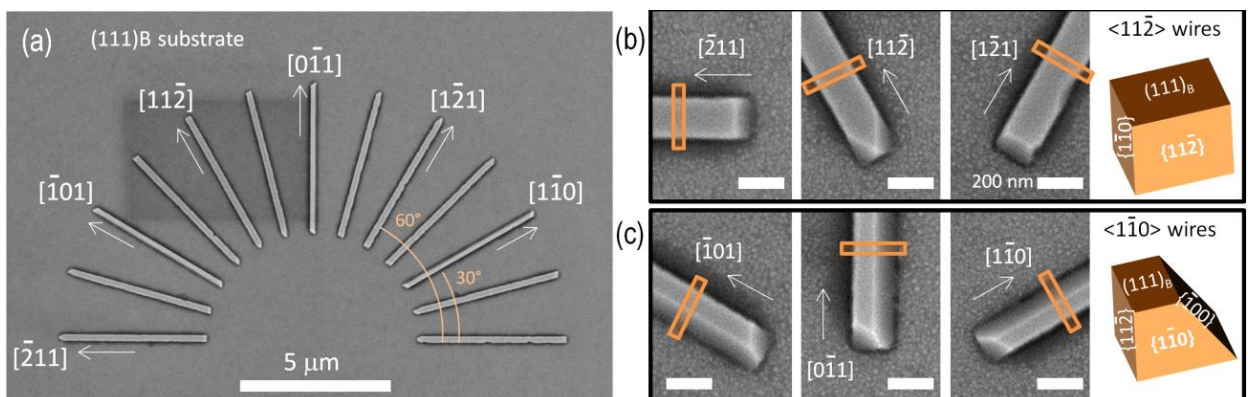

FIG 2. InAs NWs grown on an InP(111B substrate. (a) An SEM image of InAs NWs in various ridge directions. Well-defined facets are formed in particular ridge directions: (b) <11-2> and (c) <1-10>. Scale bars: 200 nm.

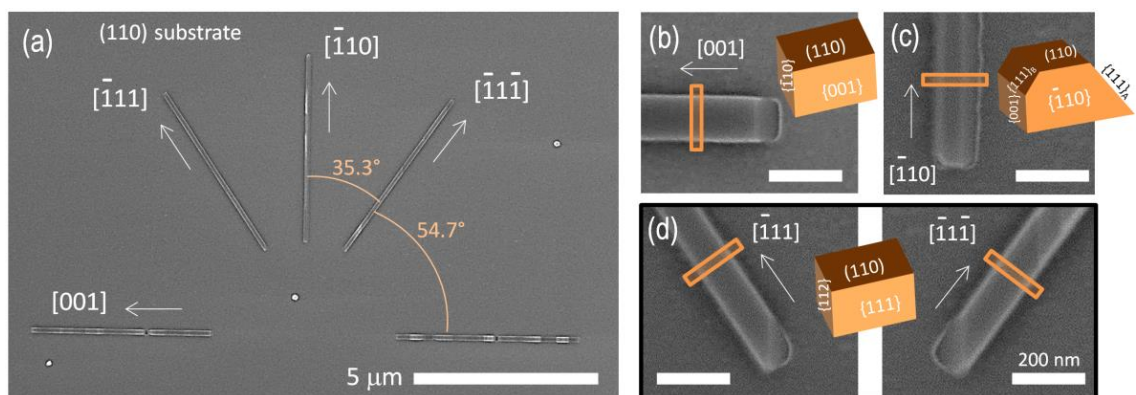

FIG 3. InAs NWs grown on an InP(110) substrate. (a) An SEM image of the InAs NWs in various ridge directions. Well-defined facets are formed in particular ridge directions: (b) [001], (c) [-110], and (d) <111>. Scale bars: 200 nm.

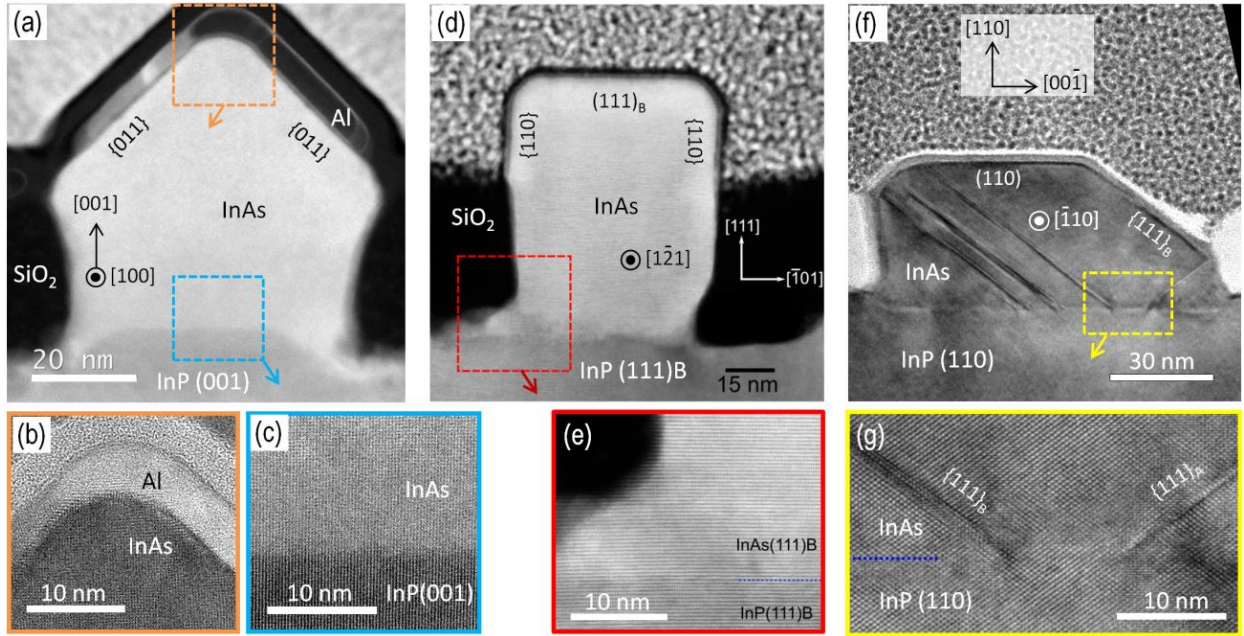

FIG 4. STEM/TEM images of NW cross-sections perpendicular to the ridge direction. (a) A HAADF-STEM image of a NW along the [100] direction grown on an InP(001) substrate exhibits slanted {011} facets. The cross-section is viewed down the [100] zone axis (along the length of the NW). Zoom-in images reveal atomically sharp interfaces (b) between InAs and Al [orange dashed box in (a)] and (c) between InAs and InP [blue dashed box in (a)]. (d) A NW along the [1-21] direction grown on an InP(111) substrate exhibits flat {-101} sidewalls and (111)$_B$ top facet in its rectangular cross-section, with a slight deviation as it grows through the SiO$_2$ mask (region of dark contrast). (e) A higher magnification image of the interface between the InP(111)B substrate and the InAs shows a continuous atomic lattice without visible extended defects, indicative of an epitaxial growth. (f) A conventional TEM image of a NW along [1-10] direction grown on an InP(110) substrate shows slanted {111} facets and stacking faults parallel to {111} planes. (g) A zoom-in image of the interface between the InP(110) substrate and InAs reveals that the stacking faults in the InAs NW originate at the interface.

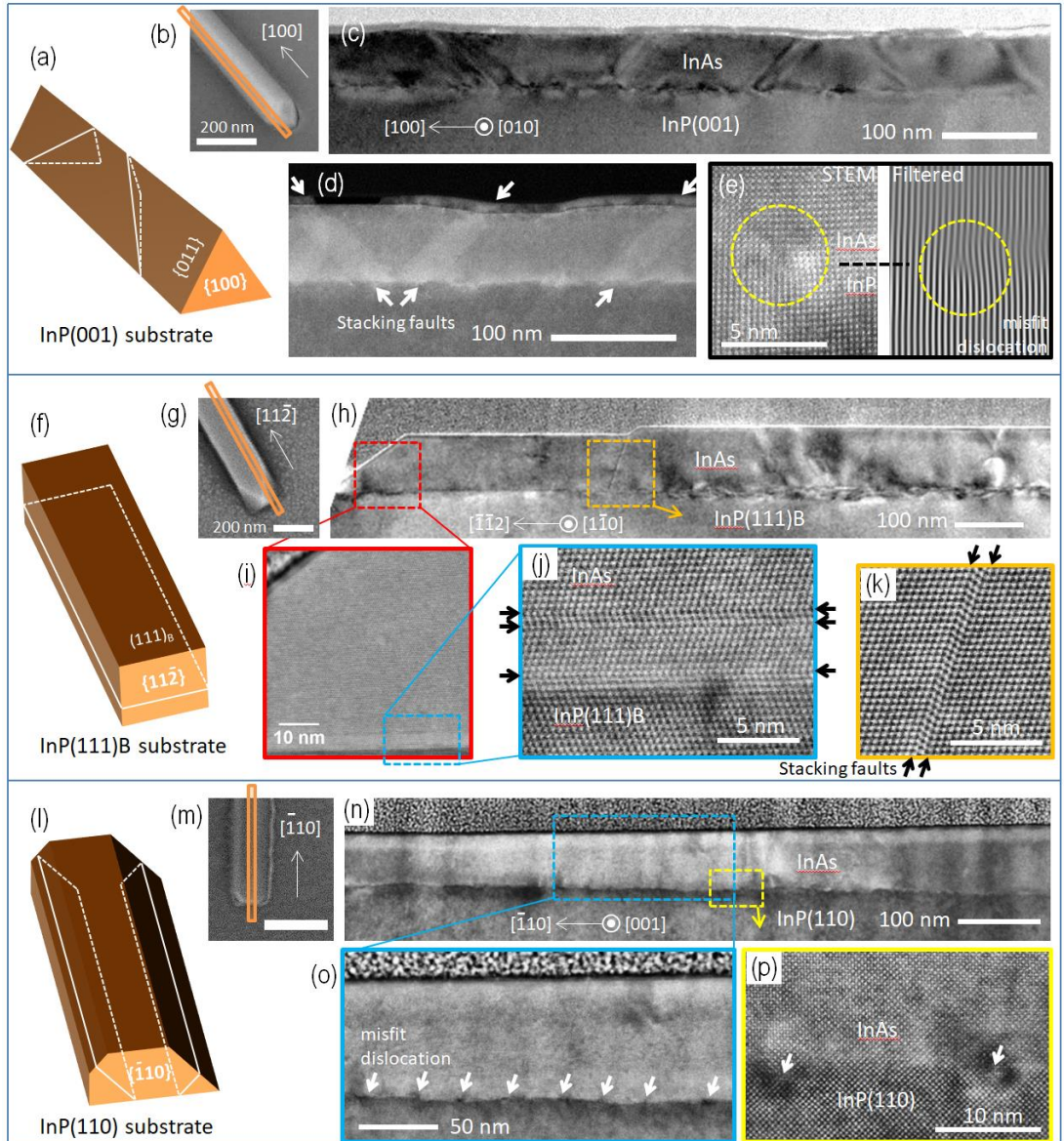

FIG 5. STEM/TEM images of NW cross-sections along the ridge direction. (a) A schematic and (b) an SEM image (b) of a NW along [100] direction grown on an InP(001) substrate. White lines in the schematic represent {111} planes in the NW, and the orange rectangle in the SEM image represents the position of the prepared cross-section. A conventional TEM image in (c) and an HAADF-STEM in (d) image reveal the inclined {111} stacking fault planes across the NW, indicated by white arrows in (d). (e) A misfit dislocation at the interface between InP and InAs is seen in an HAADF-STEM image and a filtered image. (f) A schematic and (g) an SEM image of a NW along [11-2] direction grown on an InP(111)B substrate. A conventional TEM image of the cross-section of the NW. Zoom-in images of the stacking faults near (i, j) the interface between the InP(111)B substrate and InAs and (k) a grain boundary. Black arrows represent the stacking fault planes. (l) A schematic and (m) an SEM image of a NW along [-110] direction grown on an InP(110) substrate. (n) A HAADF-STEM image of the cross-section of the NW. (o, p) Zoom-in images show misfit dislocations at the interface between the InP(110) substrate and InAs.

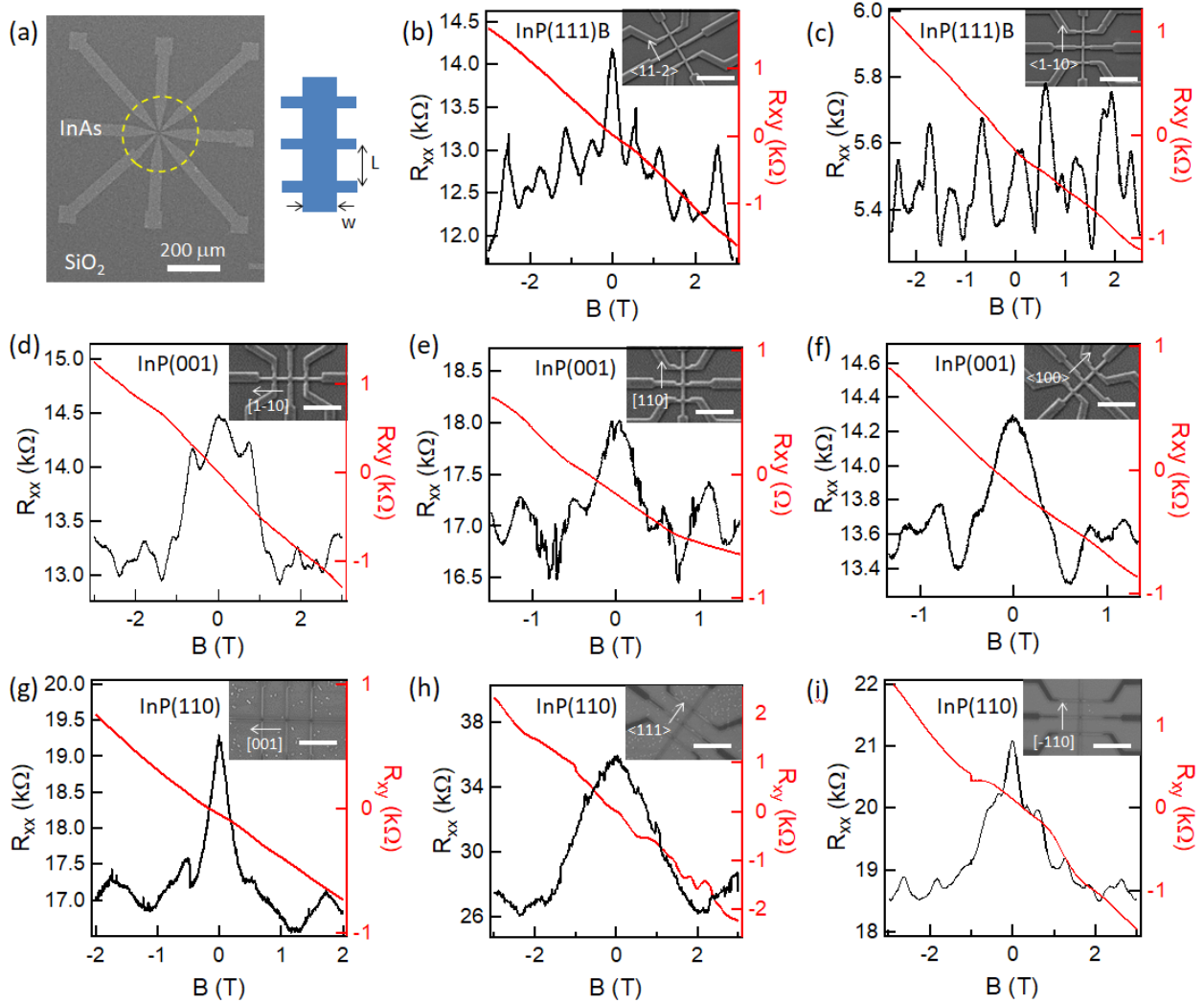

FIG 6. Hall measurements using a Hall bar geometry. (a) A low-magnification SEM image shows contact pads extended from a Hall bar, in the yellow dashed circle, illustrated as a schematic. Longitudinal resistance (black lines) and Hall resistance (red lines) were measured using Hall bar channels with various substrate orientations and channel directions: (b) along <11-2> direction and (c) <1-10> direction grown on an InP(111)B substrate, (d) along [1-10] direction, (e) [110] direction, and (f) <100> direction grown on an InP(001) substrate, and (g) along [001] direction, (h) <111> direction, and (i) [-110] direction grown on an InP(110) substrate. Insets: SEM images of measured Hall bars (scale bar: 2 μm). All measurements were carried out at 2 K.

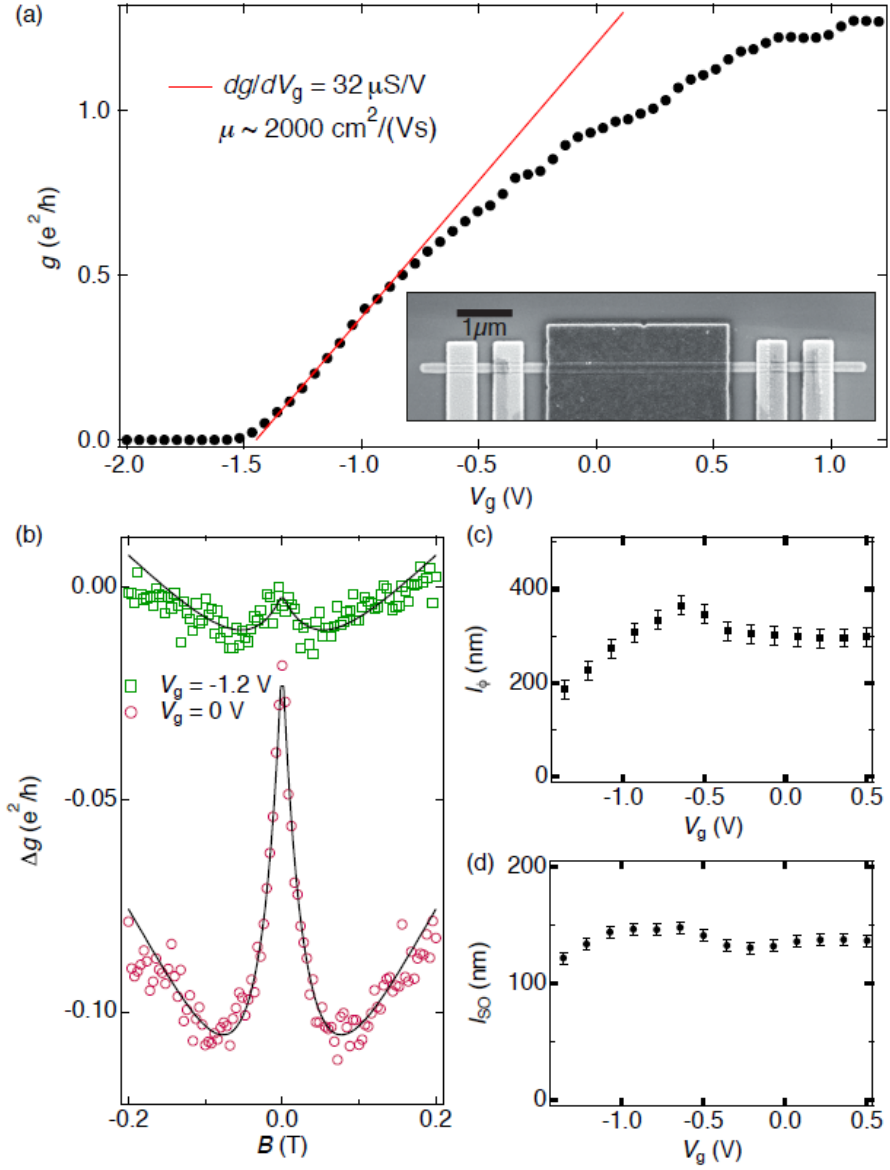

FIG 7. Transport characterization of field effect mobility and WAL. (a) Differential conductance of a selectively grown InAs NW as a function of gate voltage $V_g$. The solid red line is a linear fit to extract the peak in transconductance $g_t$. The inset shows an SEM image of the field effect mobility device. (b) Magnetoconductance $\Delta g$, offset to zero at zero field for two different gate voltages. The two curves are vertically offset. The WAL peaks are fitted with functions described in the main text (black lines). (c, d) Extracted phase coherence length $l_\phi$ (c) and spin-orbit length $l_{SO}$ (d) from WAL fits as a function of $V_g$.

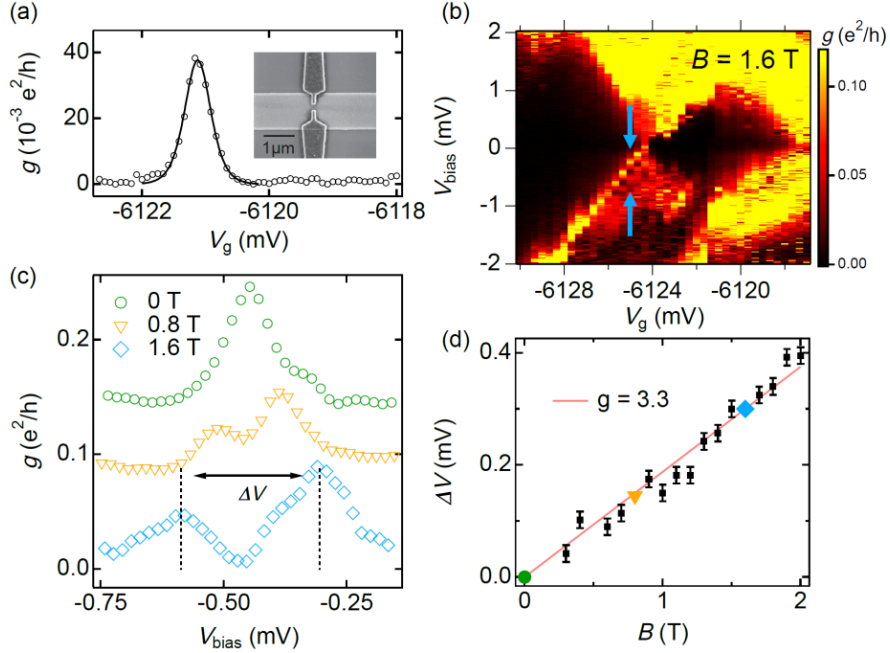

FIG 8. Measurement of the electron g-factor. (a) Coulomb peak close to pinch off in the split-gate geometry. Inset: an SEM image of the split-gate device. (b) The differential conductance $g$ is plotted as a function of $V_{bias}$ and $V_g$ in the Coulomb blockade regime at an in-plane field $B = 1.6$ T. (c) Cuts along $V_{bias}$ at a fixed $V_g$ for different magnetic fields. The curves are shifted horizontally and vertically for clarity. (d) The extracted $\Delta V$ for different in-plane fields with a linear fit to the data (red line).

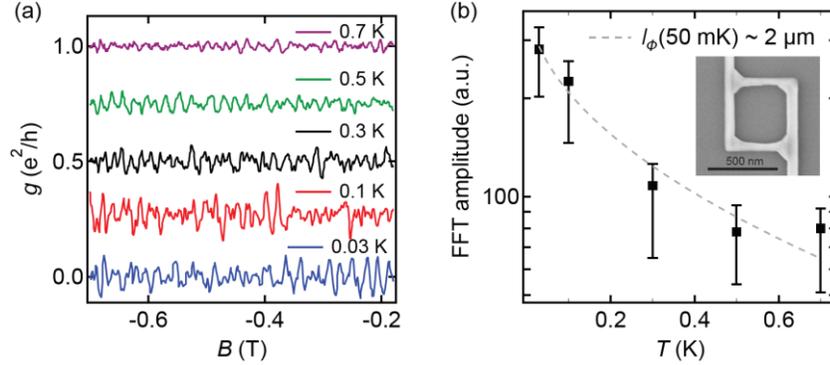

FIG 9. Measurement of the electron phase coherence length $l_\phi$ in an AB geometry. (a) Magnetoconductance with subtracted background as a function of out-of-plane magnetic field $B$ for different temperatures. Curves are vertically offset for clarity. (b) Integrated FFT of the oscillations plotted as a function of temperature. The fit (dashed grey line) corresponds to the amplitude decay within the diffusive regime with $l_\phi \sim 2$ μm at 50 mK. Inset: an SEM image of a grown AB loop.